\documentclass[aps,prd,superscriptaddress,reprint,floatfix,twocolumn]{revtex4-2}

\usepackage{amsmath, amsthm, amssymb}
\usepackage{mathtools} 
\usepackage[plainpages=false, colorlinks=true, anchorcolor=blue, linkcolor=blue, citecolor=blue, bookmarks=false]{hyperref}
\usepackage{slashed}
\usepackage{bbold}
\usepackage[utf8]{inputenc}
\usepackage{empheq}
\usepackage{graphicx}
\usepackage{siunitx}
\usepackage{xcolor} 
\definecolor{kb}{rgb}{1.0, 0.55, 0.0}

\newcommand{\rms}{r_\text{rms}}

\newcommand{\Li}[2]{\mathrm{Li}_{#1}\left(#2\right)}

\begin{document}
\title{Self-consistent bounds on Beyond the Standard Model bosons from spectroscopy of muonic atoms with magic nuclei}

\author{K. A. Beyer}
\email[Author to whom correspondence should be addressed. ]{ }
\affiliation{Max-Planck-Institut f\"ur Kernphysik, Saupfercheckweg 1, 69117 Heidelberg, Germany}

\author{N. S. Oreshkina}
\affiliation{Max-Planck-Institut f\"ur Kernphysik, Saupfercheckweg 1, 69117 Heidelberg, Germany}

\begin{abstract}
Spectroscopy of muonic atoms is, to date, the most accurate technique to extract parameters of the nuclear charge density. The same reasons for their heightened sensitivity to nuclear parameters, a large overlap of the muonic wavefunction with the nucleus, makes them attractive systems to test Beyond the Standard Model (BSM) Physics. This raises concerns of self-consistency as the same data are used to, first, extract nuclear parameters, and second, check the consistency with BSM models. We combine the two steps and self-consistently extract the nuclear and BSM parameters. We show that the data are consistent with vanishing BSM coupling and extract robust exclusion bounds. We further note that the nuclear parameters change under the influence of those BSM couplings on the parameter fits and compare with the fit solely based on quantum electrodynamics~(QED).
\end{abstract}
\maketitle


\paragraph*{Introduction:} Muonic atom spectroscopy is a well established method for the extraction of nuclear properties like the root-mean-square (rms) charge radius \cite{Wheeler1949,Feinberg:1963xj,BorieRinker1982}. The bound muon's wavefunction, because of its' increased mass compared to electrons, has a much larger overlap with the nucleus, and therefore, is much more sensitive to nuclear structure. Extraction of nuclear structure parameters relies on a fit of state-of-the-art QED calculations to experimental data \cite{Nortershauser:2023jwa}. 
However, spectroscopic measurements of $\mu-^{208}$Pb \cite{Yamazaki:1979tf,Bergem:1988zz} revealed a striking discrepancy with the prediction of the theory; the same problem exists in $\mu-^{90}$Zr \cite{Phan:1985em} and $\mu-^{120}$Sn \cite{Piller:1990zza}. As a result of this puzzle great effort has been going into improving the QED calculations since then. Historically, the largest contribution to the theoretical uncertainty, and thus hypothesized to be responsible for the discrepancy, came from nuclear polarisation which was investigated recently in~\cite{Valuev:2022tau} and found to be insufficient to reconcile the discrepancy. Further QED effects, such as muonic self-energy correction and relativistic recoil effect were improved in \cite{Oreshkina:2022mrk} and \cite{Yerokhin_murec_PhysRevA.108.052824}, respectively. The progress in theory was recently met with a full re-evaluation of the experimental data which resulted in a new fit for those elements and the resolution of the anomaly in $^{90}$Zr \cite{Pb_PhysRevLett.135.163002,inPrep}.

The spectra of the magic $\mu-^{90}$Zr, $\mu-^{120}$Sn, and  $\mu-^{208}$Pb are of special interest because the nuclear properties required for the QED calculation are fully encapsulated by the nuclear charge density $\rho_C(\mathbf{r})$. Having a nuclear ground state spin of $0$ there are no effects related to the nuclear spin and the nuclear charge density is spherically symmetric $\rho_C(\mathbf{r}) = \rho_C(r)$, making the spectrum relatively simple compared to other muonic atoms~\cite{BorieRinker1982,Michel:2018mfp}. In principle it is possible to extract the radial shape of the charge distribution from electron scattering experiments, however, in reality the finite range of momentum transfer which can be probed imply rather large uncertainties on the shape \cite{Dreher:1974pqw}. In the following we will assume a normalised Fermi-distribution for the nuclear charge density
\begin{equation}
\label{Eq:FermiDistribution}
    \rho_C(\mathbf{r})=\frac{1}{-8\pi a^3\Li{3}{-\exp\left(\frac{c}{a}\right)}} \frac{1}{1+\exp\left(\frac{r-c}{a}\right)}
\end{equation}
where $\text{Li}_n(z)=\sum_{k=1}^\infty \frac{z^k}{k^n}$ is the polylogarithm function. Parameters $c$ and $a$ are free parameters to be determined by a fit to spectral data. These fix the root-mean-square charge radius which is defined as
\begin{equation}
    \rms^2 = \int d^3r r^2\rho_C(\mathbf{r}) = 12 a^2 \frac{\Li{5}{-\exp\left(\frac{c}{a}\right)}}{\Li{3}{-\exp\left(\frac{c}{a}\right)}}.
\end{equation}

The combination of high precision and a sizeable overlap between the muonic and nuclear wavefunctions are perfect conditions to test Beyond the Standard Model (BSM) physics. The Standard Model of Particle Physics, despite its many successes, is known to be incomplete. Perhaps the most famous shortcoming is the dark matter content of the Universe but other phenomena like the baryon asymmetry are not explained within the SM~\cite{ParticleDataGroup:2024cfk}. A plethora of theories which extend the SM to address these shortcomings exist, however, the parameter space which needs to be tested is vast. Searching for BSM Physics at the high-precision frontier traditionally involves  a specific BSM model to calculate the effect on some observable which is then compared to measurements. In general this comparison requires knowledge of QED parameters, in the case at hand e.g. the rms charge radius, which were found from a separate fit to, in most cases, the \emph{same experimental data}: here the spectrum of $\mu-^{90}$Zr, $\mu-^{120}$Sn, or $\mu-^{208}$Pb \cite{Beyer:2023bwd}. This raises concerns of self-consistency \cite{Delaunay:2022grr}. Therefore, in this paper, we aim to extract self-consistent bounds on several phenomenological BSM models and, at the same time, investigate their influence on the QED parameter extraction by fitting the data to ``QED + BSM'' directly.

\begin{figure*}[t]
    \centering
    \includegraphics[width=0.80\textwidth]{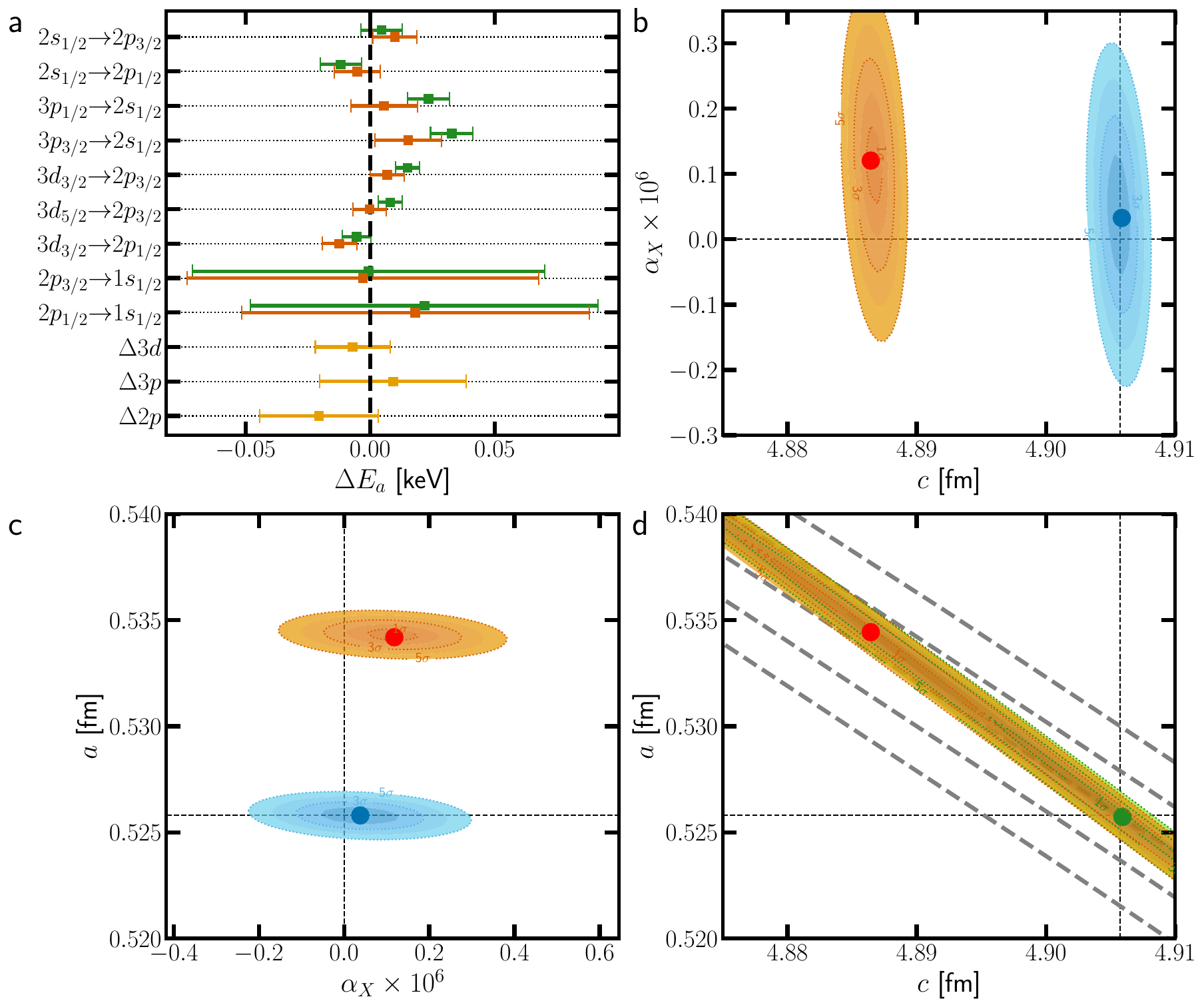}
    \caption{A plot of the $\mu-^{90}$Zr fit described in the text. Here we show results for a massless scalar boson $X$. a) The figure shows the resulting energy residuals $\Delta E_a^\mathrm{Pb}$ after the fit in orange. In green we show the previous fit for $\alpha_X = 0$. b), c), and d) show the likelihood contours for the three free parameters and the dotted lines indicate the best-fit parameters for the ´QED only' fit. The orange contours correspond to cuts through the best-fit parameters of the ´QED+BSM' fit. For convenience we show the same cuts through the BSM fit likelihood surface at the best-fit QED only parameters in light blue. This highlights the consistency of those parameters with the fit despite the deceptively far orange region. The dotted lines indicate those best-fit QED only parameters. In figure d) the black solid lines highlight the constant rms charge radius curves and the green ellipse shows the likelihood contour for the QED only case.}
    \label{fig:Zr_Fit_scalar}
\end{figure*}

\paragraph*{BSM contribution:} The BSM models we consider here are purely phenomenological; for specific models we refer the reader to the vast literature available elsewhere. Adding a boson which couples to muons and at least one of the nucleons to the standard model results in an additional force acting on the bound muon \cite{Moody:1984ba} and therefore affects the atomic spectrum. The Lagrangian density defining the coupling of such a hypothetical new Boson $X$ with SM fermions $f$ reads
\begin{equation}
    \label{Eq:Coupling_Lagrangian}
    \mathcal{L}_\varphi \supset g_f^\varphi \varphi(x)\Bar{f}(x)f(x),\qquad  \mathcal{L}_Z \supset g_f^Z Z_\nu(x)\Bar{f}(x)\gamma^\nu f(x)
\end{equation}
where $\varphi(x)$ and $Z_\nu(x)$ are scalar and vector boson wavefunctions, respectively, and $g_f$ are the coupling constants. In the present paper we restrict ourselves to scalars and vectors because of the vanishing nuclear ground state spins; the coupling to pseudoscalars and axial vectors, which couple to the spin, is greatly suppressed. Note that throughout this paper we reserve the label `$\mu$' to denote muonic quantities rather than spacetime indices.

We adopt a classical, non-relativistic description of the nucleus in which it is fully described by its, in the present case spherically symmetric, density $\rho_n(r)$. Thence, as a result of the addition of the new boson $X$, the nucleus gives rise to a BSM potential
\begin{equation}
    V_X^\mathrm{BSM}(\mathbf{r}) = -\xi_X N_n\alpha_X\int\rho_n(r)\frac{\exp\left(-m_X\lvert\mathbf{r}-\mathbf{r'}\rvert\right)}{\lvert\mathbf{r}-\mathbf{r'}\rvert} d^3r',
\end{equation}
where $\alpha_X = g_\mu^X g_n^X / 4\pi$ parametrises the strength of the coupling between muon and nucleon mediated by the new boson $X$, $m_X$ is the mass of the boson, and $\xi_\varphi = \gamma^0$ and $\xi_Z = -1$. $N_n$ is the number of nucleons $n$ in the nucleus. Note that in case $X$ couples to protons and neutrons, the coupling strength $\alpha_X$ need not be the same for both nucleons. The same is true for the densities of protons and neutrons, however, because we will find that $\alpha_X\ll 1$ we approximate all distributions by the nuclear charge distribution \cite{BABrown_1979}. 

\paragraph*{Fit to data:} In principle the BSM potential must be included in Dirac's equation to solve for the muon potential self-consistently. However, because $\alpha_X/\alpha_\mathrm{QED}\ll 1$ we restrict our analysis to lowest order in the BSM coupling. Thence, we find the binding energy of state $i$
\begin{equation}
\label{Eq:BSM_State_Correction}
    E_i [c,a,\alpha_X,m_X] = E_i^\mathrm{QED}[c,a] + \Delta E_i^\mathrm{BSM} [\alpha_X,m_X],
\end{equation}
with
\begin{equation}
    \Delta E_i^\mathrm{BSM}[\alpha_X,m_X] = \int d^3 x \mu^\dagger_i(\mathbf{x}) V_X^\mathrm{BSM}(\mathbf{x})\mu_i(\mathbf{x}).
\end{equation}
Here we have indicated the dependence of the free parameters explicitly. In general the BSM correction does depend on the free parameters $c$ and $a$, however, we note that any small change in those parameters is a higher order effect \cite{Barrett:1970wpc,Ford:1969rzz}. Thus, when evaluating the BSM correction, we will utilise muonic wavefunctions corresponding to the pure QED fit in \cite{inPrep}. This simplification is further validated by the smallness of the difference in the parameters $c$ and $a$ between the `QED' and `QED + BSM' fits.

We treat the QED contribution to the binding energy in Eq. \eqref{Eq:BSM_State_Correction} exactly as in Refs. \cite{Pb_PhysRevLett.135.163002,inPrep}. We refer the reader to these papers for details on the QED part of the procedure and will only summarise key aspects. Finite nuclear size, Uehling \cite{Uehling:1935uj}, Wichmann-Kroll \cite{Wichmann:1956zz}, and Kallen-Sabry \cite{Kallen:1955fb} are included on the level of the potential and thus to all orders. Note that the effects themselves are the first contributions in a perturbation series of the full vacuum polarisation contribution. Nuclear polarisation \cite{Valuev:2022tau}, muon self-energy \cite{Oreshkina:2022mrk}, and nuclear recoil~\cite{Yerokhin_murec_PhysRevA.108.052824} are included on the level of the energy and therefore to the lowest order in $\alpha$ only. Note that all effects are calculated in the Furry picture, thus, to all-orders in~$\alpha Z$.

We will focus on the spectrum of muonic $^{90}$Zr because its precision and fit to the theory prediction is by far the best. This makes the interpretation of the bounds rather straightforward because we can be confident that the system is well understood. We will also comment on the same procedure for muonic $^{208}$Pb, however, because of the poorer quality fit and concerns about the dependence of the spectrum on the shape of the nuclear charge distribution, less confidence can be placed in the extracted bounds. At the risk of stating the obvious, we stress that despite the bounds being extracted \emph{self-consistently} the reliability directly depends on the completeness of the SM part of the muonic spectrum. The same is true for the spectrum of muonic $^{120}$Sn which has two sets of experimental data, one older one with significantly larger uncertainties \cite{ENGFER1974509}, and one newer one with an insufficient number of resolved transitions for a self-consistent BSM fit \cite{Piller:1990zza}.

\paragraph*{Zr and Sn:} The result of the fitting to the spectrum of $^{90}$Zr in the limit of $m_X\rightarrow 0$ for the scalar case is shown in Fig. \ref{fig:Zr_Fit_scalar} and the resulting parameters are summarised in Table \ref{table:Results}. The BSM coupling is fully consistent with $\alpha_X = 0$ for light bosons. Here the QED parameters change within their error bars only and the rms charge radius is mostly unaffected. The figure shows the spectral fit with and without BSM coupling in the upper left panel. It can be seen that the fit has not meaningfully changed. The remaining panels depict various cuts through the likelihood surface of the ´QED + BSM' fit. For a fixed boson mass $m_X$ this is a $3$ dimensional object and we show two cuts, through the planes of best-fit parameters including BSM coupling (in orange) and the best-fit parameters for the QED only fit (in blue). In green we reproduced the QED only fit of \cite{inPrep}. It can be seen that the new fit is perfectly consistent with a vanishing BSM coupling and reproduces the same rms charge radius within uncertainties.

\begin{figure*}[th!]
    \centering
    \includegraphics[width=0.85\textwidth]{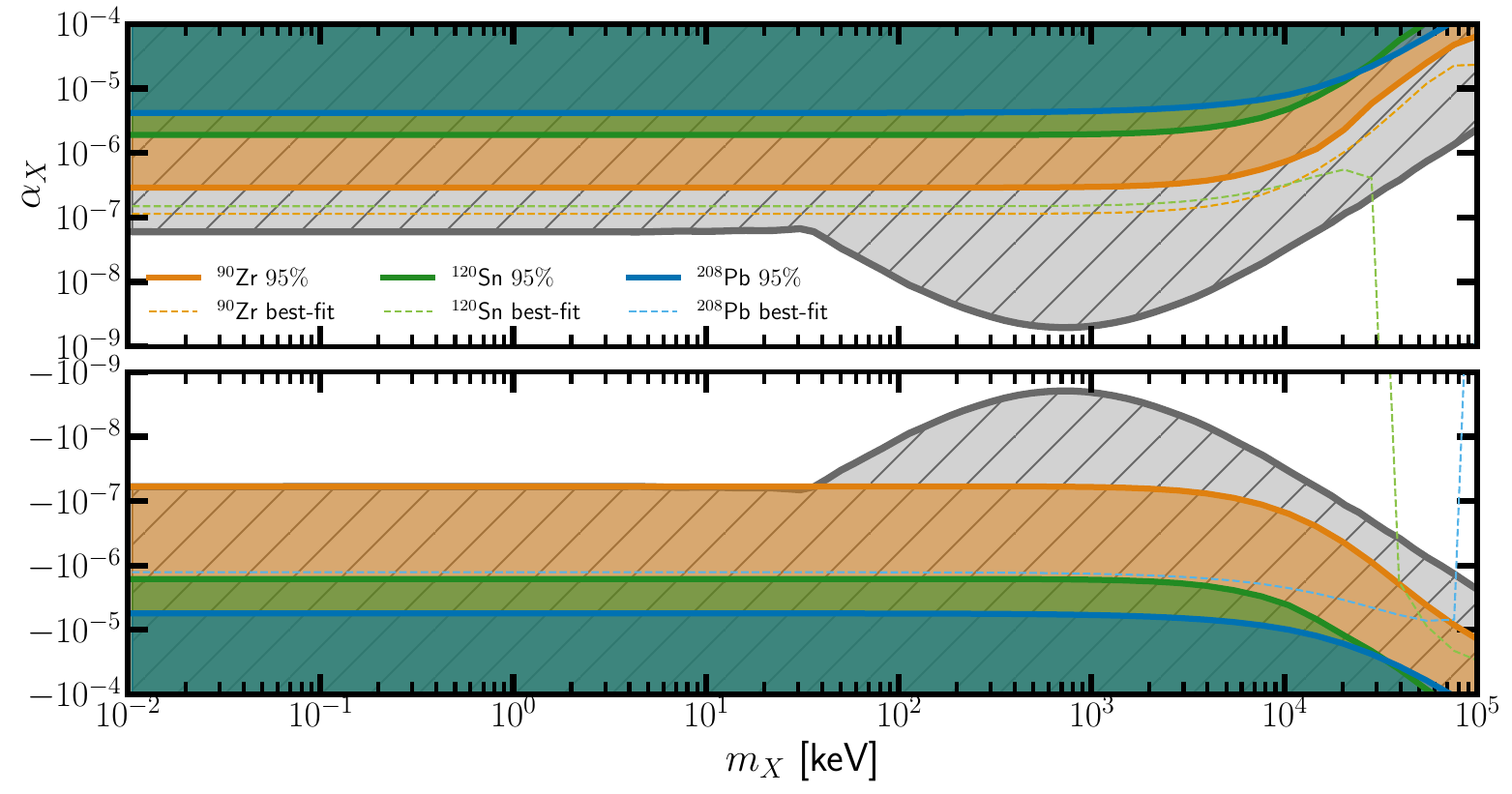}
    \caption{The plot shows the parameter space for a new scalar particle. The gray region shows previous exclusions coming from Lamb shift measurements of light muonic atoms and the transition $3d_{5/2}-2p_{3/2}$ in $\mu-^{24}$Mg and $\mu-^{28}$Si \cite{Beltrami:1985dc}. In orange we depict the self-consistent bounds coming from $\mu-^{90}$Zr spectroscopy, those from $\mu-^{120}$Sn in green, and $\mu-^{208}$Pb in blue. The dashed lines indicate the best-fit BSM coupling in the same colours as the corresponding exclusion regions.}
    \label{fig:Zr_Scalar_Excl}
\end{figure*}

The massless limit is applicable for masses $m_X a_0^\mu \ll 1 $ where $a_0^\mu$ is the muonic Bohr radius of the relevant states. Above these masses, the exponential suppression reduces the influence of the BSM potential. We have performed separate fits for a range of masses $m_X\neq 0$ and compiled the resulting exclusion bounds on the couplings to scalars in Fig.~\ref{fig:Zr_Scalar_Excl}. We do not show the relevant bounds to vectors, as they are, on a log-log plot, indistinguishable from the sign flipped scalar bounds. To find true exclusion bounds we picked coordinates in the mass $m_X$ and coupling $\alpha_X$ space to compare fits of the ´QED+BSM' model to the original ´QED-only' fit. Any point for which the latter is preferred over the former by a p-value more than $95\%$ we consider excluded. Along the exclusion bounds obtained here we evaluated the change in best-fit rms charge radius and compared it to the uncertainty of the original fit. The rms charge radius does vary significantly, around a factor $5$ over the uncertainty for lower masses and up to a factor $30$ for higher masses. This, however, must be taken with a grain of salt as the ´QED+BSM' model has double the number of free parameters and, especially close to the exclusion boundary, is not preferred over the original fit. Additionally those extreme varyations are only observed in parameter space already excluded by previous measurements. When following the lower bounds, where our bounds slightly surpass previous bounds, the rms charge radius only changes within it's uncertainty. Therefore, the variation can not simply be interpreted as an uncertainty of the rms charge radius due to BSM effects. Nevertheless, it is important to keep in mind that ´QED+BSM' models with good fit can have significantly different rms charge radii.

For heavier bosons the fit converges towards a non-zero BSM coupling as can be seen in Fig. \ref{fig:Zr_Scalar_Excl}. The figure shows the best-fit BSM coupling as a light, dashed line and the $95\%$ confidence exclusion bounds as shaded regions. It must be stressed that the quality of the fit, as measured by the $\chi^2$, does only improve by about a factor 2, hardly being statistically significant: The ´QED+BSM' fit is not statistically significantly preferred over the ´QED-only' fit. Furthermore, the preferred BSM coupling falls within the previous exclusion bounds of Ref. \cite{Beltrami:1985dc}. We therefore conclude that this result is not statistically significant and must be regarded as a statistical fluctuation. 

\begin{table*}
\centering
\bigskip
\begin{tabular}{c || c | c | c  | c  | c  }
& QED & \multicolumn{4}{ c }{QED + BSM} \\ \hline
  &  & \multicolumn{2}{ c |}{scalar}  & \multicolumn{2}{ c }{vector}\\\hline
 $m_X$  & --- & $0$ & $0.5 m_\mu$ & $0$ & $0.5 m_\mu$ \\\hline
 c [fm] & $4.8944(78)$ & $4.887(13)$ & $4.829(33)$ & $4.888(13)$ & $4.844(33)$ \\
 a [fm] & $0.5309(35)$ & $0.534(6)$ & $0.563(15)$ & $0.534(6)$ & $0.555(15)$ \\
 $\alpha_X$ & $0$ & $1.1(7)\times 10^{-7}$ & $1.1(4)\times 10^{-4}$ & $-1.1(7)\times 10^{-7}$ & $-8(4)\times 10^{-5} $ \\
 $\rms$ [fm] & $4.2740(7)$ & $4.2747(11)$ & $4.2856(50)$ & $4.2746(11)$ & $4.2829(50)$
\end{tabular}
\caption{The table summarises the best-fit values for a QED only and self consistent fits to QED + scalar and QED + vector to the spectrum of $^{90}$Zr. In both cases, negligible boson mass and large boson mass, the rms charge radius is unaffected.} 
\label{table:Results}
\end{table*}

The bounds we obtain are slightly weaker than the bounds from Lamb shift measurements of light muonic atoms, see \cite{Beltrami:1985dc}. However, they are extracted self-consistently. To compare these with the alternative, naively extracted bounds, we performed a fit of the BSM coupling to the residuals of the QED fit only. Such a fit would, incorrectly, produce bounds which are better than the self-consistent bounds by up to a factor $6.5$ for low masses and $30$ for higher masses. It must be stressed, though, that these `improved' bounds are incorrect.

\begin{figure}[t]
    \centering
    \includegraphics[width=0.48\textwidth]{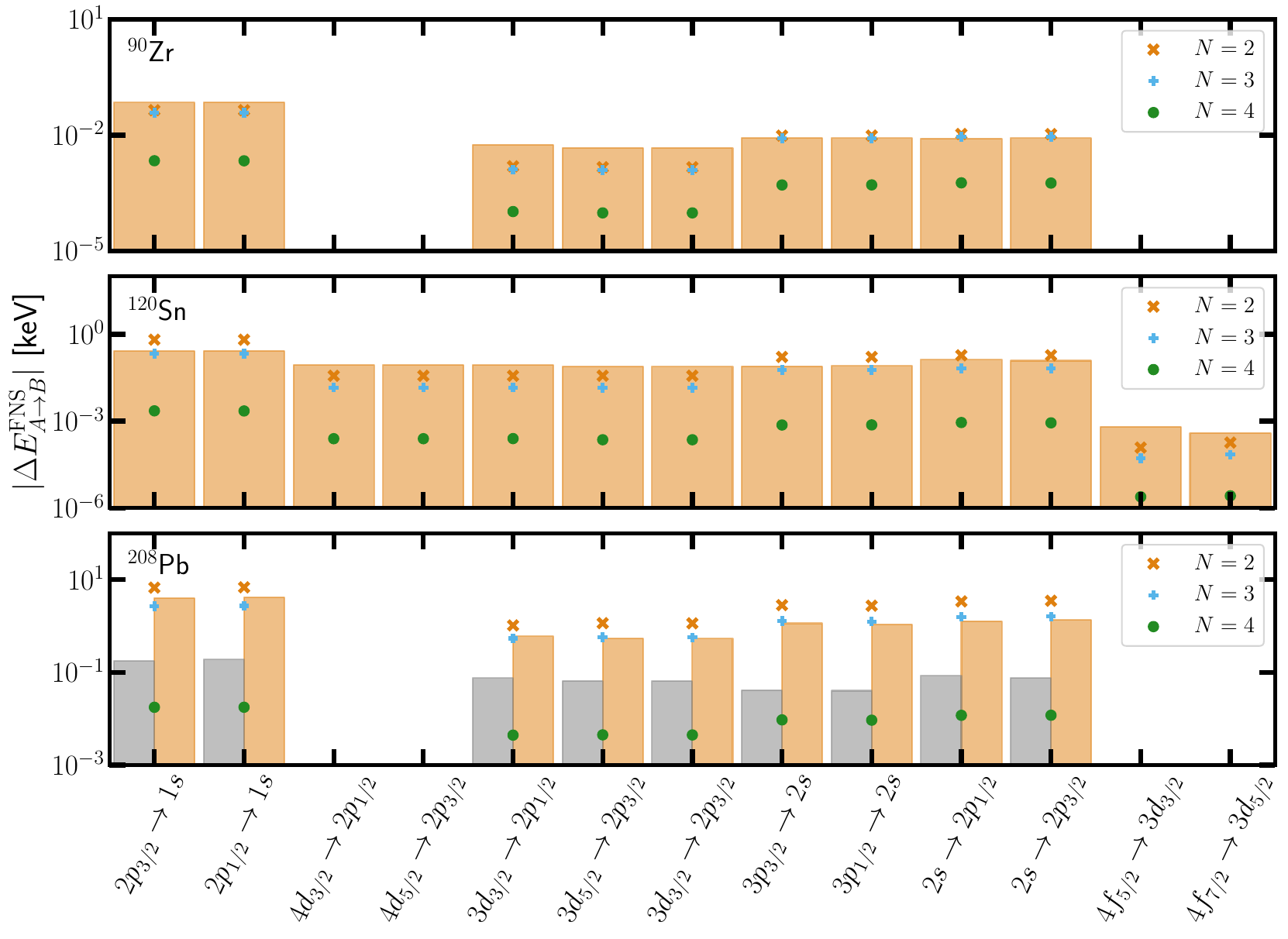}
    \caption{This plot shows the energy difference \eqref{Eq:Erg_diff} resulting from a change in nuclear charge density distribution. The first $N$ moments are fixed to coincide. As can be seen, the more moments are fixed, the smaller the energy difference will be. Further, the orange shaded areas indicate the uncertainties utilised in the fit. In the lowest plot the fit for Pb is depicted where the level of uncertainty, depicted in gray, is too small for a nuclear charge distribution with only 2 free parameters. Therefore the energy difference for $N=2$ is taken as a proxy for the shape uncertainty and added to the Pb-fit. The result of this procedure is the orange region. Note that this does affect the best-fit values, however they agree within their error bars.}
    \label{fig:moment_scaling_pb}
\end{figure}

\paragraph*{Pb:} The same analysis can be performed for the spectral data of muonic $^{208}$Pb. Here, however, more care must be taken. The ´QED-only' fit to the data is poor ($\chi^2/\text{DoF} \sim 9.5$) because of a significant theory uncertainty originating from the shape of the nuclear charge density. This uncertainty is difficult to quantify as a systematic scan of the space of all possible charge distributions is not possible. An order of magnitude however can be estimated by generating a family of distributions, Gaussians in this case, with an increasing number of free parameters $N$:
\begin{equation}
    \rho_\mathrm{C}^\mathrm{Gauss}(r) = \frac{1}{\lceil N/2\rceil}\sum_{i=1}^{\lceil N/2\rceil}\frac{1}{(2\pi \sigma_i^2)^{3/2}} \exp\left(-\frac{\left(r-\delta_i\right)^2}{2\sigma_i^2}\right).
\end{equation}
It is then possible to match the first $N$ even moments of the distribution (the odd ones vanish because of the spherical symmetry) to the 2pF distribution used in the fit
\begin{align}
    \langle r^{2n}\rangle \equiv \int r^{2n} \rho_\mathrm{C}^{2pF}(\mathbf{r}) d^3r &= \int r^{2n} \rho_\mathrm{C}^\text{Gauss}(\mathbf{r}) d^3r \nonumber\\
    &\forall \, n =1,2,...,N.
\end{align}
The first order energy difference caused by the different charge distributions is then simply
\begin{align}
\label{Eq:Erg_diff}
    \Delta& E_{A\rightarrow B}^\mathrm{FNS} = \\
    &4\pi Z \alpha \int_0^\infty dr r^2 \Delta\rho(r) \int_0^\infty dr' \frac{\lvert\Psi_A^\mathrm{2pF}(r')\rvert^2 - \lvert\Psi_B^\mathrm{2pF}(r')\rvert^2}{\mathrm{max}(r,r')} ,\nonumber
\end{align}
where $\delta \rho(r)$ is the difference between the Gaussian and the 2pF distributions and $\Psi_A^\mathrm{2pF}$ is the wavefunction of the bound muon for a 2pF charge distribution. As can be seen in Fig. \ref{fig:moment_scaling_pb} the energy difference only falls below the level of uncertainty, as indicated by the orange shading, when $N=4$. The 2pF distribution which was fitted however is $N=2$ and therefore the shape uncertainty must be included. Note that the same is not necessarily true for $^{90}$Zr, and $^{120}$Sn where $N=2$ already falls close to the uncertainty band.

We then repeat the analysis of Ref. \cite{Pb_PhysRevLett.135.163002} including the energy difference for $N=2$ as a proxy for the shape uncertainty. The covariance can easily be obtained by taking the errors to be uncorrelated in the state basis and therefore the only correlation comes from changing to a basis of transitions. We would like to point out that this is a rough estimate, nevertheless it gives an idea of the order of magnitude of the error associated to the shape. Performing this fit, the best-fit parameters for $^{208}$Pb then change to $c = 6.65(4)$fm, $a = 0.52(2)$fm which corresponds to $\rms = 5.5044(38)$fm. The quality of the fit is $\chi^2/\mathrm{DoF} = 0.11$. With this better fit of the data we can now attempt to extract BSM bounds. Those bounds are presented in Fig. \ref{fig:Zr_Scalar_Excl} by the blue shaded regions. It comes to no surprise, given the large uncertainty originating from the shape of the nuclear charge distribution, that the bounds are generally weaker than those from $^{90}$Zr. 

Once again, the fit results in a non-zero value for the BSM coupling, as indicated by the dashed blue curve, however, in the case of $^{208}$Pb $\alpha_X = 0$ is consistent within a $1\sigma$ confidence interval throughout most of the mass space. The quality of the fit improves only marginally, indicating that the ´QED+BSM' model is not statistically preferred over the ´QED-only' fit. Furthermore, it is true again that the rms charge radius is unaffected by the BSM boson.

\paragraph*{Conclusion:} On its face we have shown that the spectral data of $\mu-^{90}$Zr are consistent with a vanishing BSM coupling and extracted appropriate bounds on the parameter space, confirming the complimentary bounds obtained from spectroscopic measurements of lighter muonic atoms. The present bounds are extracted \emph{self-consistently}, by allowing the BSM effects to influence the QED parameters extracted from the fit. Thus, the bounds are conservative as this procedure generally results in weaker bounds. Nevertheless, our bounds are slightly better than previous bounds in some parameter space. This study is of further interest as we have shown that the nuclear parameters extracted from a pure QED fit are, depending on the size of the effect, affected by the inclusion of BSM physics, further highlighting the need for \emph{self-consistent} studies. The latter is of special interest as nuclear parameters extracted from muonic spectroscopy are frequently used in precision QED tests. To limit the influence of BSM effects on precision measurements of, e.g. rms charge radii, it is therefore important to combine multiple measurements in which the BSM effects either cancel or scale differently.

\bibliography{Bibliography}
\end{document}